\documentclass{mem}
\usepackage{natbib}\usepackage{txfonts}\usepackage{balance}
\usepackage{graphicx}
\usepackage[a4paper]{hyperref}
\idline{75}{282}
\begin{document}
\def\teff{$T\rm_{eff }$}
\def\kms{$\mathrm {km s}^{-1}$}

\title{
The Araucaria Project
}

   \subtitle{}

\author{
G. \, Pietrzy{\'n}ski\inst{1,2} 
\and W. \, Gieren\inst{1}
          }

  \offprints{G. Pietrzy{\'n}ski}

\institute{
Departamento de Fisica, Universidad de Concepci{\'o}n, 
Casilla 160-C, Concepci{\'o}n, Chile
\and
Warsaw University Observatory,
Aleje Ujazdowskie 4,
00-478, Warszawa, Poland
\email{(pietrzyn,wgieren)@astro-udec.cl}
}

\authorrunning{Pietrzy{\'n}ski and Gieren }

\titlerunning{The Araucaria Project}

\abstract{Results from a long-term observational project 
called the Araucaria Project are presented. Based on Wide Field 
optical monitoring of 8 nearby galaxies, covering a large range 
of metallicities,  more than 500 Cepheids 
and a few hundred Blue Supergiant candidates  were identified.
From the analysis of Cepheid P-L relations of outstanding quality derived from 
our data we conclude that the slope of these relations in the I band
and  Wesenheit index are not dependent on metallicity. Comparing the I-band
magnitudes of Cepheids of a period of ten days, as computed from our P-L
relations, to the I-band magnitudes of the tip of the RGB, which is widely
believed to be independent of population effects, we cannot see 
any obvious dependence of the zero point of the I-band P-L relation
on metallicity. A preliminary analysis of 
IR follow-up observations of sub-samples of the identified Cepheids
in various galaxies of the project 
show that the distances obtained from these data are systematically
shorter by about of 0.1 mag than those derived from the optical photometry.
It is likely that this effect can be attributed to the internal
reddening in the program galaxies. The selected Blue Supergiant candidates were observed 
spectroscopically with 8m-class telescopes to determine their element abundances,
and their luminosities from the Flux-weighted Gravity-Luminosity Relationship.
Results on this aspect of the Araucaria Project are presented in the review 
of Kudritzki presented  during this conference.
}
\maketitle{}

\section{Introduction}
The Araucaria Project was started some four years ago at the University of 
Concepci{\'o}n (Gieren et al. 2001), with the main goal to study  how 
population effects such as age and/or metallicity can affect 
the distance determination with the main stellar distance indicators.
In the course of the Araucaria Project we are mainly interested in
Cepheids, Blue Supergiants, TRGB, red clump stars, and RR Lyrae stars.
Eleven galaxies, each one containing most of the mentioned "standard candles"
and  having widely different environments
were selected, and then observed with a variety of telescopes and 
instruments. The observations have resulted in a huge database of very 
accurate measurements, very well suited for calibrating the dependence of the
brightness of the studied distance indicators on environmental 
properties, and therefore improving the whole distance scale in the 
Universe. Some preliminary results from Araucaria were presented 
by Pietrzy{\'n}ski and Gieren (2003). In this invited talk, we will  
discuss some more recent results which focus on the usefulness of Cepheid
variables as distance indicators.

\section{Observations}
The first necessary step which needed to be carried out was an extensive 
multi-epoch Wide Field optical survey for distance indicators in our target galaxies. For this 
purpose, we used the 1.3-m Warsaw telescope at the LCO, the 2.2-m MPI telescope at 
La Silla, and the 4-m Blanco telescope at the CTIO. All of these telescopes 
are equipped with Wide Field imagers (8k x 8k mosaics) having 
large fields of view (about 35 x 35 arcmin), combined with a very good spatial
resolution, allowing us to observe the whole galaxy in one shot. The
observations were typically conducted over 1-2 years  through V and I 
band filters. Between 60 and 200 V and I band images were collected for 
each program galaxy.

These data were very useful for selecting fields rich in
Cepheids  and Blue Supergiants for the follow-up  IR and spectroscopic
observations obtained with the ESO VLT telescopes located at Cerro
Paranal, and the Magellan telescopes at LCO. 

\section{Results}

\subsection{Optical survey}

In Table 1 we summarize the results for the Cepheid surveys in our target
galaxies. The columns contain the 
names of the observed galaxies, the number of discovered Cepheids, the
number of observed epochs, the range of Cepheid periods found, 
and the current status of the observational 
campaign (on-going or finished).

\begin{table*}
\caption{Results from the optical survey}
\label{abun}
\begin{center}
\begin{tabular}{lccccccc}
\hline
\\
Galaxy  & N of Cepheids & N of observations &  period range & status \\
\hline
\\
NGC 300  & 129 &  80 & 115 - 5 & finished \\
NGC 6822 & 116 &  77 & 124 - 1.7 & finished \\
WLM      &  30 & 150 &  54 - 3.5 & finished \\
NGC 3109 &  90 &  85 &  40 - 3.5 & finished \\
Phoenix  &   0 &  50 &    ----   & finished \\
NGC 55   &  80 &  70 &  121 - 6  & finished \\
NGC 247  &  50 &  84 & 114  - 11 & on-going \\
NGC 7793 &  -- &  60 &  ----     & on-going \\
\hline
\end{tabular}
\end{center}
\end{table*}

The Cepheid P-L relations in  the I band and the Wesenheit (Wi)
index for NGC 6822 and NGC 300 are shown in Fig. 1.
In Figure 2, the slopes of the P-L relations in I and Wi 
are plotted versus metallicity for  the LMC, SMC, IC 1613 
(Udalski et al. 1999, Udalski et al. 2001), Milky Way (Gieren et 
al. 2005b), NGC 300 (Gieren et al 2004), NGC 6822  (Pietrzynski 
et al. 2004), NGC 3109 and WLM (Pietrzynski et al. in preparation).
As one can see, the slopes of the P-L relations in both bands do not 
depend on metallicity, within the sensitivity of our current measurements. 
While a careful study of the possible dependence of the 
zero points of the Cepheid  P-L relations on metallicity must 
await the final analysis  of all our data, our preliminary
results indicate that the effect is very small, if at all present
(Pietrzynski et al. 2004). The catalogs of Cepheids
together with the  distance determinations for NGC
300 and NGC 6822 based on the optical data were already reported ( Pietrzy{\'n}ski et al. 2002,
Pietrzy{\'n}ski et al. 2004, Gieren et al. 2005). The papers 
with the corresponding results for the other galaxies of the Araucaria Project
will follow soon.

\subsection{Infrared follow-up} 
There are several important advantages of using infrared observations 
in studying Cepheids. The most important one is that the extinction 
in K band is about an order of magnitude smaller as compared to the V band. 
Moreover, the slope of the P-L relation in the infrared is steeper than in the optical,
 and accurate mean magnitudes of Cepheids in JHK bands can be derived from just 
one observation (Soszynski, Gieren and Pietrzynski 2005). Typical 
infrared P-L relations obtained for Cepheids in our galaxies are shown 
in Fig. 3. The slopes of the P-L relation in J and K bands in NGC 300 and IC 1613 were 
found to be almost identical to those for LMC Cepheids (Persson et 
al. 2004). Having at hand distance moduli derived in optical and
infrared bands, it is possible to derive accurate reddenings, 
and thus an improved distance. In two of the galaxies for which 
we finished the analysis of our IR data (NGC 300 (Gieren et al. 2005a), 
and IC 1613 (Pietrzynski et al. in preparation)), we obtained reddenings being
significantly larger than the foreground reddening in the directions to
these galaxies (0.1 mag vs 0.02 mag).This suggests that the internal 
reddening plays a very important role in distance determinations 
with Cepheids in optical bands, and that IR follow-up observations are a "must"
to determine accurate true distance moduli, properly corrected for the effects
of dust absorption internal to the host galaxies.

\begin{acknowledgements}
Support from the Chilean Center for Astrophysics FONDAP 15010003,
and the  Polish KBN grant 2P03D02123 is acknowledged.
\end{acknowledgements}

\bibliographystyle{aa}

\onecolumn

\begin{figure}[]
{
\resizebox{\hsize}{!}{\includegraphics[clip=true]{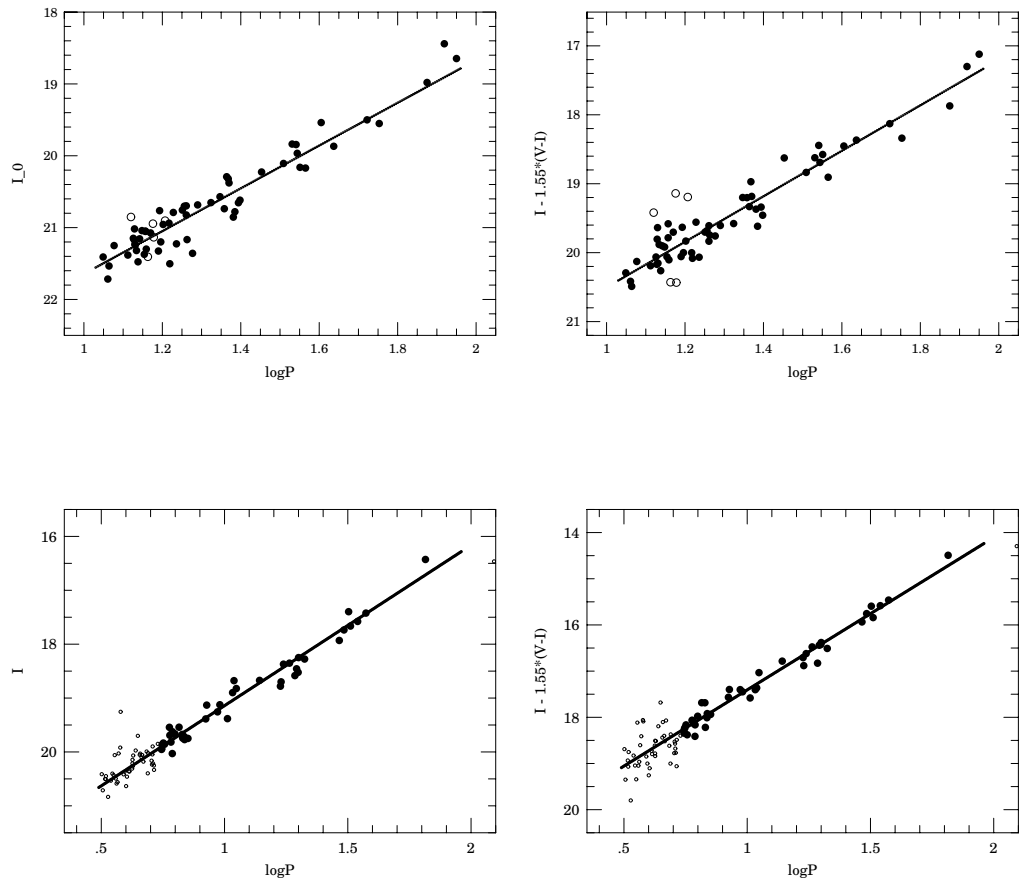}}
\caption{
Exemplary P-L relations in I and Wi bands for Cepheids in NGC 300 (upper
panels), and NGC 6822 (lower panels). }
}
\label{P-L Opt}
\end{figure}

\begin{figure}[]
{
\resizebox{\hsize}{!}{\includegraphics[clip=true]{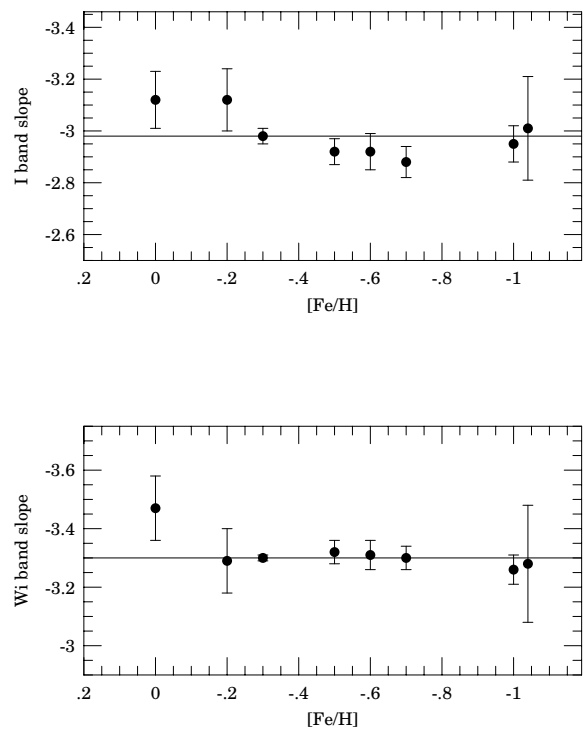}}
\caption{
Slope of the P-L relations in I and Wi bands versus metallicity 
for Cepheids in nearby galaxies. No dependence  of the slopes in both bands
on metallicity, over a broad range of metallicities, is indicated by these data. 
}
}
\label{Slopes}
\end{figure}

\begin{figure}[]
{
\resizebox{\hsize}{!}{\includegraphics[clip=true]{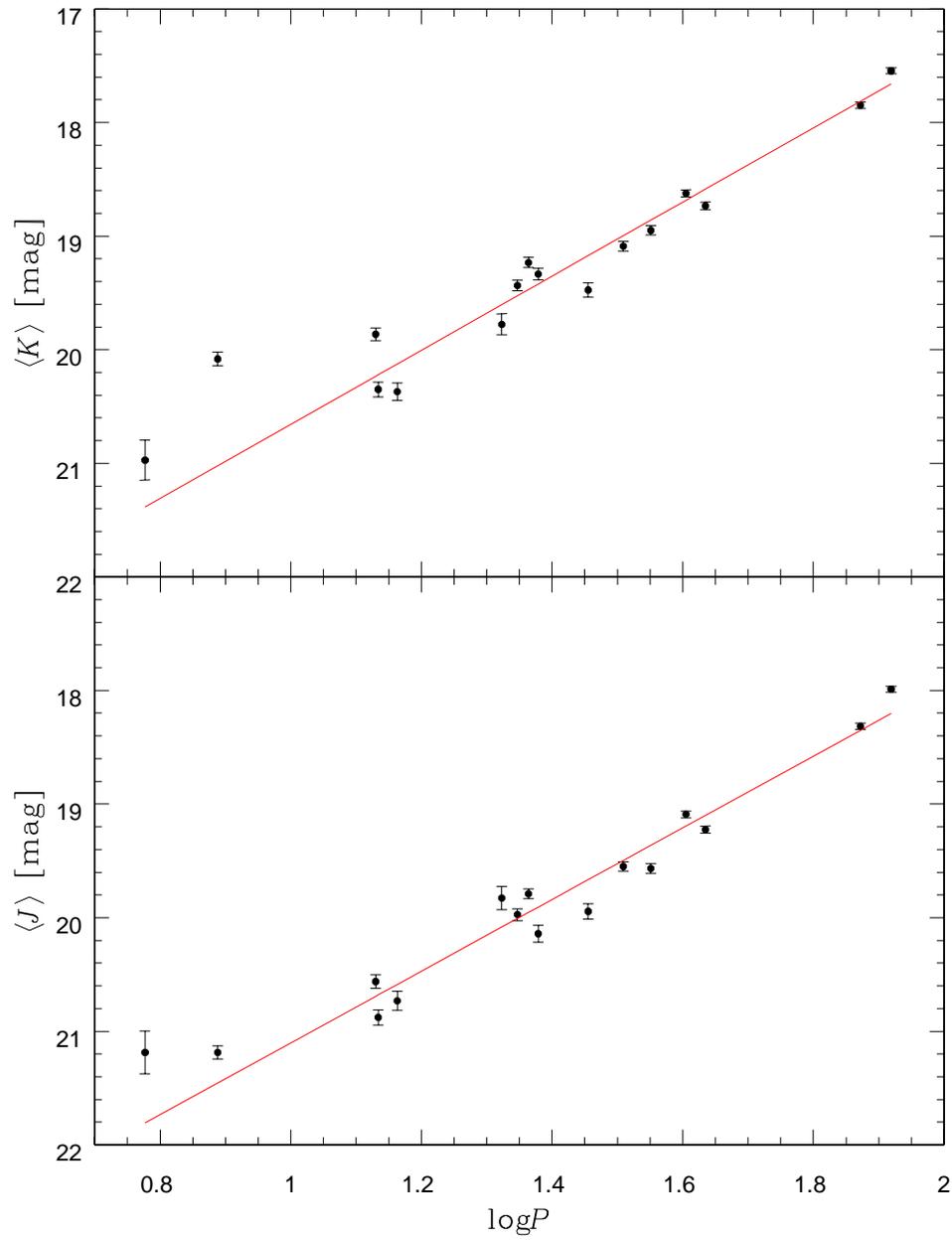}}
\caption{P-L relation in the J band (lower panel), and K band (upper panel) for Cepheids in NGC 300.}
}
\label{P-L IR}
\end{figure}

\end{document}